\documentclass[times,10pt,twocolumn]{article}
\usepackage{latex_etfa}
\usepackage{times}
\usepackage[dvipdf]{graphicx}
\usepackage[dvips]{color}

\begin{document}

\title{Energy Efficient and Reliable Wireless Sensor Networks - An Extension to IEEE~802.15.4e}

\author{Achim Berger, Markus Pichler\\
\\
{\normalsize Linz Center of Mechatronics GmbH}\\
{\normalsize Sensors and Communications}\\
{\normalsize Altenbergerstr. 69}\\
{\normalsize 4040 Linz, Austria}\\
{\normalsize achim.berger@lcm.at}\\
\and
Andreas Springer, Werner Haslmayr\\
\\
{\normalsize Johannes Kepler University Linz}\\
{\normalsize Institute for Communications Engineering}\\
{\normalsize and RF-Systems}\\
{\normalsize Altenbergerstr. 69}\\
{\normalsize 4040 Linz, Austria}\\
}

\maketitle
\thispagestyle{empty}

\begin{abstract}
Collecting sensor data in industrial environments from up to some tenth of battery powered sensor nodes with sampling rates up to $100\;\mathrm{Hz}$ requires energy aware protocols, which avoid collisions and long listening phases. The IEEE~802.15.4 standard focuses on energy aware wireless sensor networks (WSNs) and the Task Group 4e has published an amendment to fulfill up to 100 sensor value transmissions per second per sensor node (Low Latency Deterministic Network (LLDN) mode) to satisfy demands of factory automation. To improve the reliability of the data collection in the star topology of the LLDN mode, we propose a relay strategy, which can be performed within the LLDN schedule. Furthermore we propose an extension of the star topology to collect data from two-hop sensor nodes. The proposed Retransmission Mode enables power savings in the sensor node of more than $33\%$, while reducing the packet loss by up to $50\%$. To reach this performance, an optimum spatial distribution is necessary, which is discussed in detail.
\end{abstract}

\label{section_introduction}

The technological progress in wireless microelectronics offers the usage of battery powered sensor nodes with acceptable battery life times in many areas from consumer electronics to industrial applications \cite{industrial_WSN_challenges}. Miniaturized, battery powered, and self-sufficient sensor nodes are sensing physical properties and the wireless sensor network (WSN) is responsible for transmitting the sensed data to a base station for further processing. The requirements on the data traffic can be very diverse from real-time, reliability, scalability, deterministic behavior, to low-power. Each application has its individual set of requirements.

In our work we focus on WSNs for real-time data collection from battery or energy harvester powered miniaturized sensor nodes \cite{energy_harvesting_technologies} to mains powered data sinks with transmission of up to 100 sensor values per second and high reliability. The data sink (later called coordinator and the gateway to a wired industrial network) is stationary, while the sensor nodes (later called device) are distributed over an industrial plant. Most of the devices are stationary, but might also be mobile. The time varying wireless channels between the nodes and the gateway are characterized by multipath propagation due to large-scale reflectors, moving and stationary obstacles, LOS and nonLOS links, and different interferers. The limited power resources in the sensor nodes require a balance between energy awareness and retransmission attempts. The first aim of this work is an improvement of the reliability while saving energy in the sensor nodes. Therefore we introduce a (wherever applicable mains powered) relay node which supports the sensor nodes in saving energy and increasing data transmission reliability by performing retransmissions.

We decided to use the most widely adopted standard for low-rate WSN and wireless personal area networks (WPANs), IEEE~802.15.4 \cite{IEEE_STD_802.15.4}, which is optimized for energy-efficiency, low-cost implementation, and low data rate traffic. The standardized Physical Layer (PHY) is already available in a variety of transceivers from different semiconductor companies \cite{CC2520}, \cite{MRF24J40}. Two competing protocols, which use the IEEE~802.15.4 PHY, are the wireless version of the fieldbus HART, WirelessHART, and the ISA100.11a standard. Petersen et al. \cite{wirelessHART_versus_ISA100_11a} described the basics of these two protocols and discussed the pro and cons of both systems. The evaluation in \cite{wirelessHART_energy_analysis} shows a significant portion of energy consumption due to (over)listening, which can be seen as wasted energy. An improvement of the IEEE~802.15.4 Medium Access Layer (MAC) by optimized scheduling is shown in \cite{guarantee_real_time_in_802_15_4}.

The IEEE~802.15 Task Group 4e (TG4e) was intended to amend IEEE~802.15.4 to better support the industrial markets and permit compatibility with modifications being proposed within the Chinese WPAN \cite{IEEE_STD_802.15.4e-2012}. The amendment of TG4e offers new MAC-layer possibilities which are optimized amongst others for factory automation. In factory automation usually many sensor and actuator nodes (e.g. up to 100) are connected in star topology requiring deterministic data transmission at low latency and high reliability \cite{first_variant_of_IEEE_802.15.4e}, \cite{study_of_IEEE_802.15.4e}.

Low Latency Deterministic Networks (LLDN) is a specified mode in the IEEE~802.15.4e Amendment of the IEEE~802.15.4 WPAN Standard which is designed for deterministic applications in industrial environments, where frequency planning should achieve minimal interference from other RF systems. The focused applications range from monitoring purposes with short delays to closed loop control with sampling rates up to $100\;\mathrm{Hz}$. The main benefit of the LLDN mode is the possibility of periodically transmitting sensor data from an (typically battery powered) "LLDN device" to the "LLDN coordinator" in $10\;\mathrm{ms}$ cycles, which has not been possible in a beacon enabled mode of the IEEE~802.15.4 standard. Another very important feature of the IEEE~802.15.4e standard are the provided time slots for retransmission, which are fundamental for our proposed transmission modes. Our first approach is a retransmission mode which saves energy in the device by using a new relay node. In case of bad link quality between device and coordinator, characterized by a high packet error rate ($PER$), this mode is more beneficial in terms of reliability and energy consumption compared to the retransmission strategy of the LLDN mode. But the benefits can only be achieved if the position of the relay is optimized. Tripathi et al. describe similar position optimization for a base station, where individual adjustable transmission powers are assumed \cite{optimal_position_in_two_tiered_WSN}.

Our second proposal in this work targets the expansion of the coverage area by relay nodes. The network topology in the LLDN mode is limited to a star topology by definition and the covered area is restricted to a maximum diameter of two times the communication range. To expand the area of the WSN, data must be forwarded from devices outside the communication range of the base station via a relay node. We propose a way to extend the star topology to an extended star topology, so that the coordinator can collect data from so-called two-hop devices in a deterministic way, within the same superframe. A typical way of connecting two-hop nodes to a coordinator are relay strategies \cite{survey_multipoint_relay} or routing techniques in a tree, clustered tree \cite{cross_layer_topology_optimization} or mesh network topology, which usually uses more time and energy due to the required organization of the network \cite{routing_in_wsn}.

The second proposal of this work is forwarding data from a device out of communication range to the coordinator in real-time by a second type of relay node. Our understanding of real-time is to have the sensor data available at the coordinator in the same superframe, in which they are generated, i.e. in $10\;\mathrm{ms}$ periods. All proposed modifications can be performed in the standard conform LLDN mode or with just minor modifications.

The paper is organized as follows: Existing standards and relevant publications are discussed in Section~\ref{section_related_work}. Section~\ref{section_LLDN_mode} introduces the LLDN mode with its standard conform retransmission strategy. Our proposals are presented in Section~\ref{section_retransmission_mode} and \ref{section_extended_topology_mode}. The comparison and the quantification of the improvement are discussed in Section~\ref{section_analysis} and concluded in Section~\ref{section_conclusion}.

\section{Related Work}
\label{section_related_work}
WirelessHART is an already established protocol for wireless industrial sensing applications. The schedule of WirelessHART is based on the Time Synchronized Mesh Protocol (TSMP, \cite{TSMP}), which enables very flexible, reliable, and scalable wireless mesh networks. This protocol was further developed in the Time Synchronized Channel Hopping (TSCH), which is part of the IEEE~802.15.4e Amendment and improved by different scheduling and synchronization algorithms \cite{Adapt_Sync_in_802_15_4e}. The timing is organized in time slots, which are collected to so-called slotframes. A TSCH time slot typically allows one data frame transmission with the corresponding acknowledgment frame. Therefore a single slot has to be long enough to accommodate reception of CCA (Clear Channel Assessment), if enabled, followed by the switch to transmission, packet transmission itself, switch to reception, and finally acknowledgment reception, if enabled. Thus WirelessHART and most other protocols based on TSCH or TSMP use a time slot duration of $10\;\mathrm{ms}$ and so it is not applicable to perform up to 100 transmissions of sensor values per second from up to some tenth of sensor nodes.

The idea of cooperative communication is to improve reliability of a wireless link between a source and a sink node by support of a third node, a relay node. Different MAC protocol designs are discussed and compared in \cite{cooperative_diversity_in_WLAN_and_WSN} and \cite{cooperative_protocols_survey}. The achievable diversity gain, which can be the improvement of reliability or reduction of power consumption, depends on the applied relaying strategy. The applicability and efficiency of relaying strategies strongly depend on the cause of the transmission failure. A performance analysis regarding cooperative communication and the right selection of up to two relay nodes is done in \cite{performance_of_relay_selection}. Laneman \textit{et. al} \cite{cooperative_diversity} show the benefits of cooperative schemes against the adverse effect of channel fading. Srinivasan \textit{et. al} \cite{srinivasan_kappa_factor} \cite{srinivasan_beta_factor} presented wireless link indicators to evaluate the performance of routing and network coding protocols. Their so-called $\kappa$ factor indicates the inter-link reception correlation, while the $\beta$ factor describes the burstiness disturbance of the wireless channel. LLDN mode is assumed to be operated in a controlled RF environment \cite{IEEE_STD_802.15.4e-2012}, thus the transmission errors are mainly caused by multipath propagation, which results in a $\kappa$ factor around zero, which means independent receptions in up- and downlink, even if the probability of error is the same in both links due to the reciprocity of multipath propagation.

\section{LLDN Mode of 802.15.4e}
\label{section_LLDN_mode}
\subsection{General}
\label{subsection_LLDN_mode_general}

\begin{figure}[!h]
\centering
\includegraphics[width=8.5cm]{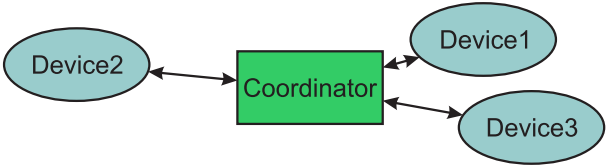}
    \caption{Example of the spatial distribution in the star network topology considered in the IEEE~802.15.4e standard.}
    \label{fig_spat_distr_standard_star}
\end{figure}

The LLDN mode of IEEE~802.15.4 uses three states, the "Discovery State", the "Configuration State", and the "Online State". After powering up the nodes of the network, they are in the "Discovery State", where the coordinator scans the environment for applicants to the network and waits for their response. After finishing the "Discovery State" the network nodes enter the "Configuration State", where the setup of the network is performed. The relevant state for data transfer is the "Online State", where data packet transmissions are executed in a TDMA (Time Division Multiple Access) schedule for a star network topology (Fig.~\ref{fig_spat_distr_standard_star}).

\begin{figure*}[t!]
\centering
\includegraphics[width=11.6cm]{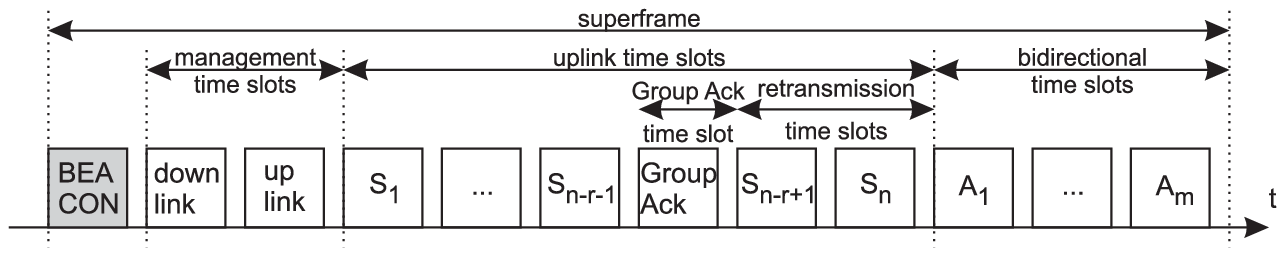}
    \caption{Superframe in the LLDN mode of the IEEE~802.15.4e standard with separate Group Acknowledgement frame \cite{IEEE_STD_802.15.4e-2012}.}
    \label{fig_802_15_4e_superframe}
\end{figure*}

The TDMA schedule in the "Online State" is organized in superframes, which are divided into a slot for a beacon, optionally time slots for management, and time slots of equal length for data transfer. The versatile configuration possibilities allow to setup a time slot as dedicated to one device or assigned to a group of devices, the communication in the slot as unidirectional or bidirectional, and as acknowledged or not \cite{IEEE_STD_802.15.4e-2012}. The LLDN superframe contains also retransmission time slots, which offer low latency retransmission attempts in the same superframe, which are the basis for our two proposals. The high degree of flexibility of the LLDN mode can also be seen in the user-defined fragmentation of downlink, uplink, and retransmission slots, as shown in Fig.~\ref{fig_802_15_4e_superframe}.

The slot configuration we want to consider in the following modes uses the feature of a separate Group Acknowledgement frame (GACK) in a dedicated uplink time slot while no management and no bidirectional time slots are used. The GACK frame includes the bit-mapped acknowledgement information, one bit for each time slot.

\subsection{Standard-conform retransmission}
\label{subsection_standard_conform_retransmission}
\begin{figure*}[!t]
\centering
\includegraphics[width=17.0cm]{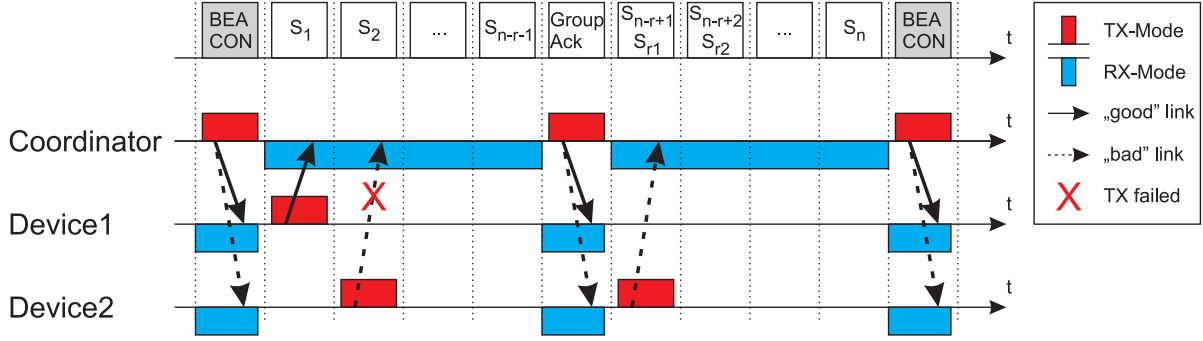}
    \caption{Example of the schedule in the standard mode (SM). Device1 transmits successfully to the coordinator, while device2 needs a retransmission.}
    \label{fig_standard_star_schedule}
\end{figure*}

The Standard Mode ("SM") considers a network topology as shown in Fig.~\ref{fig_spat_distr_standard_star} and a schedule shown in Fig.~\ref{fig_standard_star_schedule}, which represents one LLDN configuration example of IEEE~802.15.4e, where $n$ is the total number of time slots (beacon slot excluded, GACK slot included) and $r$ the number of re-transmit slots. A $10\;\mathrm{ms}$ superframe provides eight regular and eight re-transmission slots, one pair for each node, with two bytes payload ($n=17$ and $r=8$). The coordinator is the gateway and is mains powered, while the devices are battery powered and in the communication range to the coordinator.

The exemplary schedule in Fig.~\ref{fig_standard_star_schedule} shows a successful first transmission from device1 to the coordinator, while the first attempt of device2 is not successful. In this work we distinguish between "good" and "bad" links, which represent wireless links with low and high $PER$ without quantifying them. The reasons for different link qualities range from path loss to static and dynamic multipath propagation \cite{fundamentals_wireless_comm}, \cite{channel_measurements_koerber_scholl} to fast moving or rotating sensor nodes \cite{PER_for_rotating_sensor_nodes} \cite{speed_dependent_PER}. For the calculations regarding spatial distribution of link qualities in section~\ref{section_analysis}, we will use large-scale propagation models, which characterize signal strength over transmission distances from $2$ to $100\;\mathrm{m}$ \cite{RAPP}. This is a simplification compared to a realistic factory environment \cite{channel_measurement_for_802_15_4_transceiver}, but allows to derive generic results.

Due to the information transmitted in the GACK frame, device2 receives permission to transmit its packet in the first retransmission time slot $S_\mathrm{r1}$ again. The device's transceiver activities in one superframe in case of a successful transmission are one transmission and two receptions. Device2, which requires a second transmission, needs two transmissions and two receptions.

To calculate the energy consumption of the nodes, information about the quality of the wireless link is necessary. We characterize the link quality by the packet error rate ($PER$). To perform a generic analysis, we distinguish between the uplink $PER$ (from devices to the coordinator $PER_\mathrm{D2C}$) and downlink $PER$ (from coordinator to the device $PER_\mathrm{C2D}$). The probability for a retransmit of any device in SM is (when either the data packet is lost and the GACK indicates a negative acknowledge, or the GACK is not received by the device and thus initiates a retransmission)
\begin{equation}
P_\mathrm{SM,Retr} = PER_\mathrm{C2D} + PER_\mathrm{D2C} - PER_\mathrm{C2D} PER_\mathrm{D2C},
\label{equ_prob_retransmit_case1}
\end{equation}
if we assume the packet errors in up- and downlink to be independent.
Thus the energy consumption due transmission and reception in the devices in one superframe results in
\begin{eqnarray}
E_\mathrm{SM,Device} = &(1+P_\mathrm{SM,Retr}) E_\mathrm{TX,Data} + \nonumber \\ & E_\mathrm{RX,Beacon} +  E_\mathrm{RX,GACK}, \label{equ_energy_case1}
\end{eqnarray}
where $E_\mathrm{TX, Data}$, $E_\mathrm{RX, Beacon}$ and $E_\mathrm{RX, GACK}$ are the consumed energy per transmission and reception of data packet, beacon and group acknowledgment.

A second important parameter is the probability of loosing data packets, i.e. packets which are not received successfully by the coordinator, which are evaluated by the Packet Loss Rate ($PLR$). Compared to the $PER$, where each lost packet between transmitter and receiver is considered, the $PLR$ is the probability of loosing data after the retransmission attempt(s).

The data packet transmission in SM is successful, when either the first transmission attempt (prob. of $1-PER_\mathrm{D2C}$) or the retransmission is received by the coordinator (prob. of $PER_\mathrm{D2C}(1-PER_\mathrm{D2C})$). These two cases result in the $PLR$ of\\

\begin{eqnarray}
PLR_\mathrm{SM} & = & 1- [ (1-PER_\mathrm{D2C}) + \nonumber \\
& & PER_\mathrm{D2C}(1-PER_\mathrm{D2C}) ] \nonumber \\
& = & PER_\mathrm{D2C}^2.
\label{equ_lost_data_case1}
\end{eqnarray}

\section{Retransmission Mode}
\label{section_retransmission_mode}
\begin{figure}[!h]
\centering
\includegraphics[width=8.5cm]{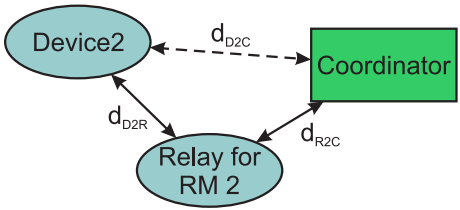}
    \caption{Example of the spatial distribution in the star network of the IEEE~802.15.4e standard with the additional relay node for RM. (Solid arrow represents "good" link; dashed arrow represents "bad" link.)}
    \label{fig_spat_distr_retransmission_proposal}
\end{figure}

To improve the reliability of the data packet transmission while staying compatible with the standard, we propose the introduction of a novel Retransmission Mode (RM) which uses a relay node.

Fig.~\ref{fig_spat_distr_retransmission_proposal} shows an example for a spatial distribution which can use the novel RM. In this configuration the LLDN device (e.g. device2 in Fig.~\ref{fig_spat_distr_retransmission_proposal}) operates as a sensor node without ACK demand and has a direct wireless link to the coordinator, probably with low reliability, i.e. high $PER$ ("bad" link). To improve the reliability of this link, we introduce a new type of node, the "relay for RM" node, which is stationary and mains powered. Relay\_for\_RM\_2 in Fig.~\ref{fig_spat_distr_retransmission_proposal} is such a node, which can communicate with device2 and the coordinator. To have the coordinator and the assigned device in its transmission range is a mandatory requirement for the relay for RM node.

\begin{figure*}[!t]
\centering
\includegraphics[width=17cm]{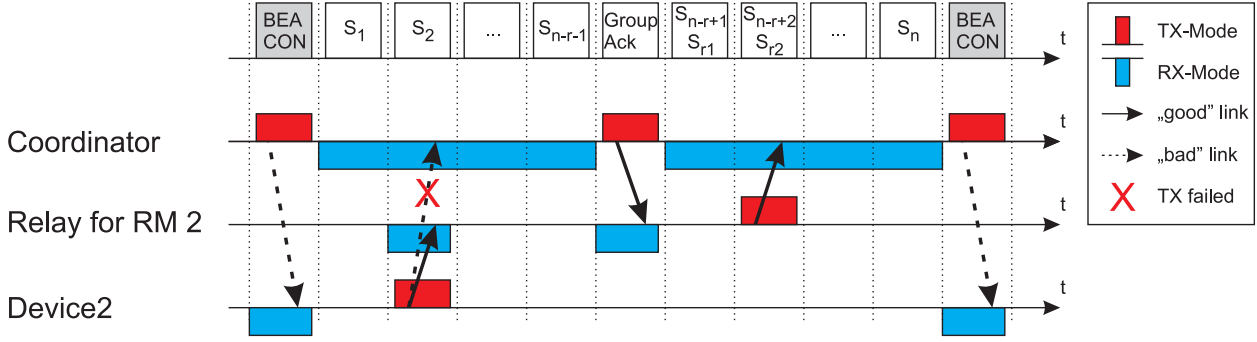}
    \caption{Example of the proposed retransmission mode (RM) with one relay node and one device}
    \label{fig_repeater_retransmission_schedule}
\end{figure*}

The task of the relay\_for\_RM\_2 node, as shown in the schedule in Fig.~\ref{fig_repeater_retransmission_schedule}, is to listen for data packets, which are sent from device2 to the coordinator and retransmit it in case of a negative acknowledgment. Therefore it has to operate in receiving mode during the regular packet transmission and the GACK transmission, and to transmit a data packet if needed. If the device is configured without ACK demand, it does not listen for GACK and thus would not disturb the retransmission attempt of the relay. So there is no need of adaptations in the device's software. In the example shown in Fig.~\ref{fig_repeater_retransmission_schedule}, the first transmission attempt of the data packet fails and the relay for RM retransmits the data packet because it received the packet successfully and knew about the failed attempt of device2 due the GACK information. Device2 has to perform only one transmission (regular transmission in the assigned time slot) and one reception (for the beacon).

The proposed relay for RM node is capable to support the retransmission for a few devices and thus should be placed (if possible) where it can establish the best connection to all associated nodes. In most applications it is predetermined where the sensor nodes have to be placed. So the positions of the relay nodes are subject to optimization. Optimization criteria are the energy consumption and the probability of loosing data packets.

The computation of the energy consumption has to be performed for the device and the relay. Like in SM the energy consumption is a function of the link quality.
The relay for RM node assumes the need for a retransmission if either the packet transmission from the device to the coordinator or the GACK transmission from the coordinator to the relay failed. Furthermore, to enable a retransmission, it is necessary, that the relay node receives the data packet from the device successfully. The probability that a retransmission of the relay to the coordinator is necessary, assuming that the retransmission slot is assigned and known a priori, is
\begin{equation}
\begin{array}{rcl}
P_\mathrm{RM,Retr} &=& (1-PER_\mathrm{D2R})\cdot [PER_\mathrm{D2C} + \\
&&PER_\mathrm{C2R} - PER_\mathrm{D2C} PER_\mathrm{C2R}].
\label{equ_prob_retransmit_case2}
\end{array}
\end{equation}

The energy consumption of a single device per superframe is constant, because it always results from only the reception of one beacon and transmission of one data packet:
\begin{equation}
E_\mathrm{RM,Device} = E_\mathrm{RX,Beacon} + E_\mathrm{TX,Data}.
\label{equ_energy_device_case2}
\end{equation}

If the relay supports only one device, it consumes energy twice per superframe for listening and $P_\mathrm{RM,Retr}$ times for retransmission of a data packet:
\begin{equation}
E_\mathrm{RM,Relay} = E_\mathrm{RX,Data} + E_\mathrm{RX,GACK} + P_\mathrm{RM,Retr} E_\mathrm{TX,Data}.
\label{equ_energy_repeater_case2}
\end{equation}

The data packet transmission in RM is successful, when either the first transmission attempt by the device (prob. of $1-PER_\mathrm{D2C}$) or the retransmission by the relay is received by the coordinator (prob. of $PER_\mathrm{D2C}(1-PER_\mathrm{D2R})(1-PER_\mathrm{R2C})$). These two cases result in the $PLR$ of
\begin{equation}
\begin{array}{rcl}
PLR_\mathrm{RM} & = & PER_\mathrm{D2C}- \\ & & PER_\mathrm{D2C}(1-PER_\mathrm{D2R})(1-PER_\mathrm{R2C}).
\label{equ_lost_data_case2}
\end{array}
\end{equation}

\section{Extended Topology Mode}
\label{section_extended_topology_mode}

The aim of the proposed Extended Topology Mode (ETM) is to extend the area covered by the WSN and enable sensor nodes, which are not in the communication range of the coordinator, to still transmit their packets to the coordinator. This task can be handled by mesh networks with routing strategies. However the required protocols usually are much more complex than in star networks and thus require increased overhead and power consumption. The advantage of our proposal is that the network organization can be performed in the standard conform simple LLDN framework which is supplemented by the "relay for ETM" node. This is a newly defined type of node different to the relay for RM of the previous subsection, but also stationary and mains powered.

\begin{figure}[!h]
\centering
\includegraphics[width=8.5cm]{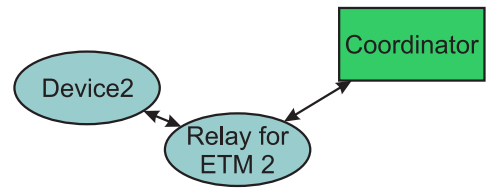}
    \caption{Example of the spatial distribution in the extended star network with the relay for ETM node for forwarding the data of the two-hop neighbor.}
    \label{fig_spat_distr_extended_topology_proposal}
\end{figure}

The proposed schedule is based on opportunistic coding, which enables forwarding of multiple packets in a single transmission. Like shown in \cite{XORs_in_the_air}, forwarding of a simple XOR conjunction of two incoming packets in a router node, enables the pairwise exchange of packets in only three transmissions instead of four. In our proposal it is necessary to exploit the limited resources available for each device in the LLDN framework. In the example shown in Fig.~\ref{fig_spat_distr_extended_topology_proposal}, device2 and coordinator do not have a direct data link between them and thus use relay\_for\_ETM\_2 to exchange the beacons and the data packets.

\begin{figure*}[!t]
\centering
\includegraphics[width=17cm]{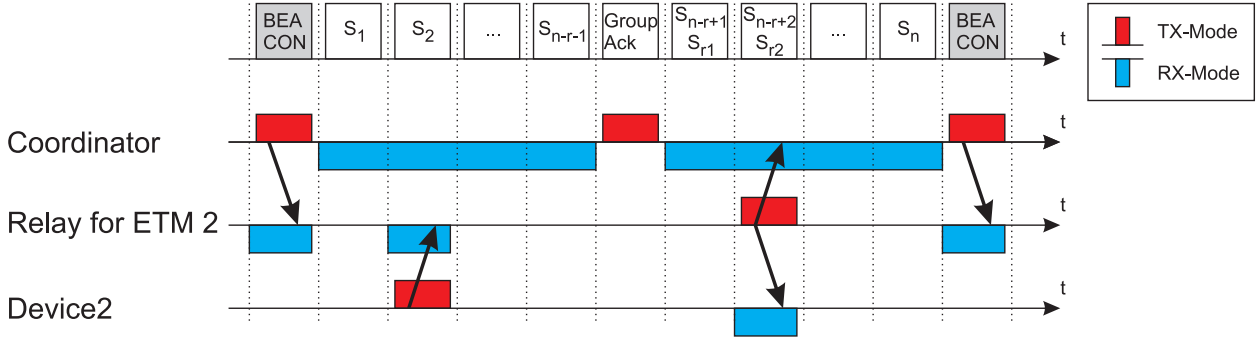}
    \caption{Example of the schedule for the proposed extended topology mode (ETM)}
    \label{fig_extended_topology_schedule}
\end{figure*}

The schedule in Fig.~\ref{fig_extended_topology_schedule} shows the packet transmission in the extended topology. The relay\_for\_ETM\_2 listens for the beacon from the coordinator and the data packet from device2 in the assigned time slots. With the assumption, that the retransmission slot is assigned and known a priori, the relay node has the possibility to use the retransmission slot for forwarding the data packet to the coordinator and in the same time slot forwarding the beacon to device2. These two packets can be encoded by XOR-ing them together to one packet. Both receivers, the coordinator and the device2, are able to decode this packet by XOR-ing it with their respective original information.

Periodical data transmission in SM leads to a worst-case delay between data generation and availability in the base station of one superframe duration, e.g. $10\;\mathrm{ms}$. This guaranteed latency increases in ETM due to forwarding the data by the relay node to one and a half superframe duration.

The reception of the data packet in the retransmission slot is part of the LLDN framework and thus without changes for the coordinator. The device requires a small modification in the timing, because the listening of the beacon is not performed at the beginning of the superframe, but in the retransmission slot. The scheduling of data packet transmission is the same like in the standard star topology. Forwarding the beacon enables the two hop device to synchronize to the LLDN network and organize its schedule regarding the beacon information, e.g. Transmission State and Configuration Sequence Number.

Similar to the RM, the placement of one or more relay nodes allows for optimization of the functionality of the network. As the energy analysis will show, the energy consumption in ETM is constant in all devices, so only the successful throughput can be optimized.

The required transmissions and receptions of the device within a superframe are a data packet transmission in the assigned time slot and the listening for the XOR-ed packet in the assigned retransmission time slot. Thus the energy consumption per superframe is constant and equals
\begin{equation}
E_\mathrm{ETM,Device} = E_\mathrm{TX,Data} + E_\mathrm{RX,XORed}.
\label{equ_energy_device_case3}
\end{equation}

The relay for ETM node performs listening for beacon, listening for data packet and transmission of the encoded (XOR-ed with beacon) data packet. The only possibility of saving energy could be skipping the encoded packet, when both reception attempts fail. In this analysis we do not consider this optimization option, because skipping of the encoded packet leads to a longer listening period in the device, because it listens until a time-out. The constant need of energy per superframe in the relay for ETM is
\begin{equation}
E_\mathrm{ETM,Relay} = E_\mathrm{RX,Beacon} + E_\mathrm{RX,Data} + E_\mathrm{TX,Data}.
\label{equ_energy_repeater_case3}
\end{equation}

Data loss occurs, when either the data packet from the device to the relay or the retransmitted data packet from the relay to the coordinator gets lost. The probability that a data packet gets lost, characterized by $PLR$, is
\begin{equation}
PLR_\mathrm{ETM} = PER_\mathrm{D2R} + PER_\mathrm{R2C} - PER_\mathrm{D2R} PER_\mathrm{R2C}.
\label{equ_lost_data_case3}
\end{equation}

In this analysis we assumed the energy consumption due to the opportunistic coding (XOR operation) to be negligible.

\section{Analysis}
\label{section_analysis}
The comparison of the regular retransmission behavior in the LLDN network Standard Mode (SM) with the proposed Retransmission Mode (RM) and Extended Topology Mode (ETM) is made based on energy consumption and on the probability of loosing data packets.  Finally the placement of the relay for RM is optimized.

\subsection{Energy Consumption}
\label{subsection_compare_energy_consumption}
The energy consumption per superframe is calculated by equation~(\ref{equ_energy_case1}) (device) for Standard Mode, equations~(\ref{equ_energy_device_case2}) (device) and (\ref{equ_energy_repeater_case2}) (relay for RM) for Retransmission Mode, and equations~(\ref{equ_energy_device_case3}) (device) and (\ref{equ_energy_repeater_case3}) (relay for ETM) for Extended Topology Mode. The energy, which is consumed during reception and transmission depends on transmission power, listening duration, packet length, and hardware specific features, like duration of PLL locking. The characterization of transceiver power consumption is described in \cite{TDMA_for_testbed}, where the values from the datasheet are verified by measurements. We used the values for a Texas Instruments CC2520 transceiver for a supply voltage of $3\;\mathrm{V}$ and a transmission power of $0\;\mathrm{dBm}$ for all nodes for the following analysis (Table~\ref{table_parameter}, \cite{TDMA_for_testbed}).

\begin{table}[h!]
\caption{Parameter used for analysis.}
\label{table_parameter}
\begin{tabular*}{8.1cm}{ll}
\hline
Parameter \hspace{4cm} & Value   \\ \hline
TX-current@0dBm ($I_\mathrm{TX,HC}$) & 25.8 mA \\
RX-current ($I_\mathrm{RX,HC}$) & 22.3 mA\\
StartUp-current ($I_\mathrm{LC}$) & 7.4 mA\\
StartUp-time ($t_\mathrm{LC}$) & 192 $\mu$s\\
Supply Voltage & 3 V \\
Data rate & 250 kbps \\
LL-Beacon packet length & 14 byte\\
LL-Data packet length & 11 byte\\
LL-GACK packet length & 12 byte\\
Number of time slots $n$ & 17 \\
Number of re-transmit slots $r$ & 8 \\
Path loss exponent $\kappa$ & 3 \\ \hline
\end{tabular*}
\end{table}

The parameters which specify the packet length are the sensor value quantization and the number of bits in the GACK. The settings in this analysis are 2~bytes for each sensor value in a LLDN network with maximum 8 devices, which lead to 1~byte of GACK information. The resulting packet lengths including overhead are 14~bytes for each beacon, 12~bytes for a GACK packet and 11~bytes for the data packets. These settings enable a superframe duration of $10\;\mathrm{ms}$ with the proposed retransmission possibility.

\begin{figure}[!h]
\centering
\includegraphics[width=8.5cm]{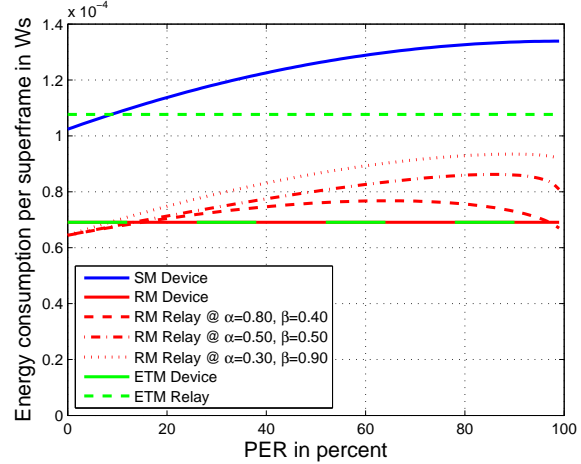}
    \caption{Comparison of the energy consumption due to transceiver activity in one superframe in the device and the relay in the three different modes.}
    \label{fig_compare_energy_per_sf}
\end{figure}

The first step for comparing the different modes is to compute the consumed energy for each transmission and reception, considering the parameters for each mode. The second step is to calculate the number of transmissions and receptions, by evaluating the retransmission possibilities of equations~(\ref{equ_prob_retransmit_case1}) and (\ref{equ_prob_retransmit_case2}). The retransmission possibilities depend on the link qualities ($PER$) between the involved nodes. In Fig.~\ref{fig_compare_energy_per_sf} the consumed energy per superframe by the device and relay nodes of the three modes as function of $PER$ are compared. This common $PER$ value represents different link qualities for each mode, which are described in detail below.

SM represents the default setup, where $PER_\mathrm{D2C}$ is valid between the device and the coordinator for the distance $d_\mathrm{D2C}$ and packet length $l_\mathrm{Data}$. The value $PER$ in Fig.~\ref{fig_compare_energy_per_sf} represents $PER_\mathrm{D2C}$ for SM and RM.
We assume that all other $PER$s, which are required for the analysis ($PER_\mathrm{D2R}$ and $PER_\mathrm{R2C}$), can be derived from $PER_\mathrm{D2C}$ --- which is assumed to be known --- simply by assuming the distances between transmitter and receiver and the transmission powers.

This calculation from a known $PER_\mathrm{1}$ to a new $PER_\mathrm{2}$ with the knowledge of the ratio of the two transmission powers ($\mathcal{E}_\mathrm{t,1}/\mathcal{E}_\mathrm{t,2}$), the ratio of the two distances ($d_1/d_2$), and the lengths of the packets ($l_1$ and $l_2$), can be done as shown in Appendix. The described calculation is based on Rayleigh fading channel model with AWGN. The probability of transmission error is determined by the signal-to-noise-ratio ($SNR$) at the receiver, which is proportional to the transmitted bit energy and inversely proportional to the distance between transmitter and receiver to the power of the path loss exponent ($SNR \sim \mathcal{E}_\mathrm{t}/d^{\kappa}$). The only assumption regarding the propagation model is the path loss exponent $\kappa$, which we assume with $\kappa=3$.

In the analysis of this subsection, all nodes use the same transmission power, consequently a transmission energy ratio of one is used ($\mathcal{E}_\mathrm{t,1}/\mathcal{E}_\mathrm{t,2} = 1$). The distances are defined by $d_\mathrm{D2R} = \alpha\cdot d_\mathrm{D2C}$ and $d_\mathrm{R2C} = \beta\cdot d_\mathrm{D2C}$, as shown in Fig.~\ref{fig_spat_distr_retransmission_proposal}. The spatial arrangement leads to the condition $\alpha + \beta \geq 1$.

To analyze the behavior of the ETM, the value $PER$ in Fig.~\ref{fig_compare_energy_per_sf} is used to describe the link qualities between device and relay for ETM and relay for ETM and coordinator. The connection between device and coordinator is by definition not considered.

In Fig.~\ref{fig_compare_energy_per_sf} we compare the energy consumption per superframe for device and relay of SM, RM, and ETM in three different spatial configurations, characterized by $\alpha$ and $\beta$. The consumed energy of the device in RM and ETM is the same and constant (green-red solid line in Fig.~\ref{fig_compare_energy_per_sf}). This value is significantly smaller (reduction by $33$ to $48\%$) than the energy consumption of the device in SM. Also the consumed energy in the relay for RM node is smaller than the energy consumption of the device in SM, when it cooperates with only one device. In that case powering by a battery is possible. If the relay forwards data from many devices, the power consumption would increase and mains powering should be considered.
The relay for ETM consumes a constant amount of energy, independent from the link quality and at $PER$ values higher than $9\%$ this consumption is smaller than that of the device in SM. As mentioned already for the relay for RM, the number of supported devices affects the energy consumption.

The main advantage of RM and ETM is the power-saving in the devices, where in RM this saving goes hand in hand with an improvement of the reliability (see subsection~\ref{subsection_compare_lost_data}), while ETM achieves a larger coverage area. The total energy consumption of device and relay in RM and ETM is higher compared to the consumption of the device in SM, but this is a moderate cost compared to the benefits.

\subsection{Lost Data Packets}
\label{subsection_compare_lost_data}
A further important advantage of the RM configuration is the reduction of lost data packets. In this subsection the Packet Loss Rate $PLR$, which depends on the spatial configuration and the $PER$, is evaluated. By $PLR$ we consider the lost packets after retransmission attempts. This is different to $PER$, where each lost packet between transmitter and receiver is considered. Equations (\ref{equ_lost_data_case1}), (\ref{equ_lost_data_case2}) and (\ref{equ_lost_data_case3}) describe the $PLR$ for the corresponding mode.

\begin{figure}[!h]
\centering
\includegraphics[width=8.5cm]{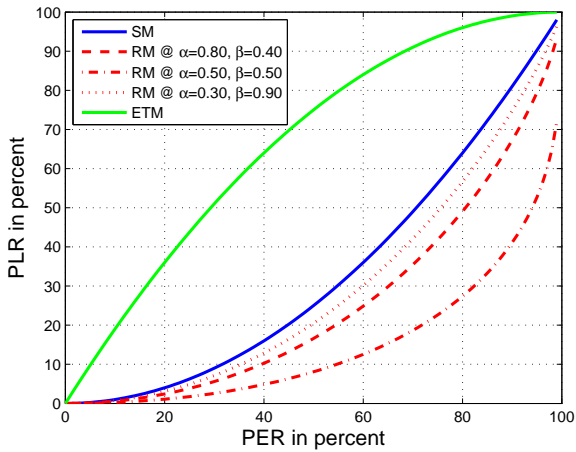}
    \caption{Comparison of the probability of loosing data packets by $PLR$.}
    \label{fig_compare_prob_lost_data}
\end{figure}

If no retransmission would be offered $PLR$ would be equal to $PER$. The improvement with only one retransmission possibility in Standard Mode is represented by the "SM" marked line in Fig.~\ref{fig_compare_prob_lost_data}.

The evaluation of the $PLR$ in the ETM, where $PER$ describes the link between device and relay as well as relay and coordinator, leads to the "ETM" marked line in Fig.~\ref{fig_compare_prob_lost_data}. Here we have to conclude that the advantage of a larger covered area has the disadvantage of a higher data loss.

The $PLR_\mathrm{RM}$ in the RM is not only a function of the $PER$, but also the placement of the relay has an influence. Fig.~\ref{fig_compare_prob_lost_data} shows that for the three placement configurations ($\alpha$ and $\beta$) already used in Fig.~\ref{fig_compare_energy_per_sf}. The intuitively best choice of relay placement in the middle between device and coordinator ($\alpha = 0.5$ and $\beta = 0.5$) leads to the best performance with a reduction of $PLR_\mathrm{RM}$ by up to $40\%$ (at $90\%$ $PER$), although the power consumption in the relay is not a maximum. The configuration $\alpha = 0.3$ and $\beta = 0.9$ has the same total transmission distance ($1.2\cdot d_\mathrm{D2C}$) like the $\alpha = 0.8$ and $\beta = 0.4$ configuration, but can not reach its performance even at a higher energy consumption. This leads to the survey about the optimum placement of the relay for RM in the following subsection.

\subsection{Spatial Configuration in RM}
\label{subsection_compare_different_spatial_configurations}
Figs.~\ref{fig_compare_energy_per_sf} and \ref{fig_compare_prob_lost_data} show that the position of the relay for RM can be optimized to achieve a network with low $PLR_\mathrm{RM}$ at low power consumption. In the analysis of the spatial configuration in RM we assume fixed positions of device and coordinator with a known $PER_\mathrm{D2C}$ at $0\;\mathrm{dBm}$ transmission power for $l_\mathrm{Data}=88\;\mathrm{bits}$ long packets and a path loss exponent of $\kappa=3$. The device is placed at $(0/0\mathrm{m})$ and the coordinator in a distance of $50\;\mathrm{m}$ at $(0/50\mathrm{m})$.
The $PLR_\mathrm{RM}$ is calculated for different positions of the relay for RM node and different transmission powers of the battery powered device $P_\mathrm{t,Device}$ ($0\;\mathrm{dBm}$, $-3\;\mathrm{dBm}$, and $-6\;\mathrm{dBm}$) with the corresponding $PER$ values and equation (\ref{equ_lost_data_case2}). The $PER$ values are calculated as described in Appendix.

\begin{figure}[!h]
\centering
\includegraphics[width=8.5cm]{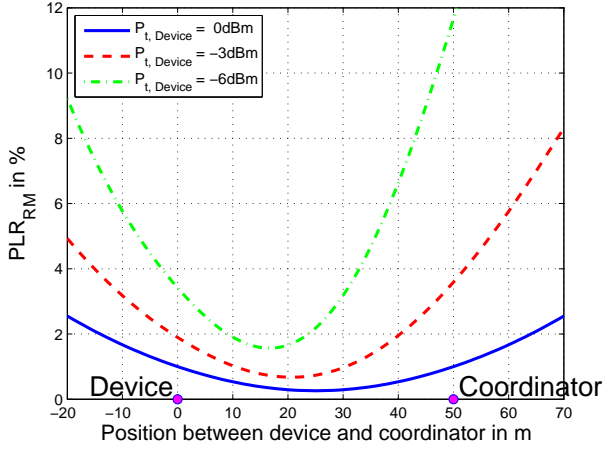}
    \caption{$PLR_\mathrm{RM}$ with different transmission power settings of the device $P_\mathrm{t, Device}$ for different positions of relay for RM on the line between device and coordinator.}
    \label{fig_prob_loosing_packets}
\end{figure}

The contour plot of Fig.~\ref{fig_prob_loosing_packets_at_one_prob} shows lines of equal $PLR_\mathrm{RM}$ in percent with different transmission powers of the device as parameter, assuming a $PER_\mathrm{D2C}$ of $10\%$. If device and relay transmit with the same power $P_\mathrm{t}$, the contour lines have their center exactly half way between device and coordinator. With decreasing $P_\mathrm{t, Device}$ the center of the contour lines moves towards the device. This can also be better seen in Fig.~\ref{fig_prob_loosing_packets}, where $PLR_\mathrm{RM}$ is shown if the relay is located on the line connecting device and coordinator.

\begin{figure}[!h]
\centering
\includegraphics[width=8.5cm]{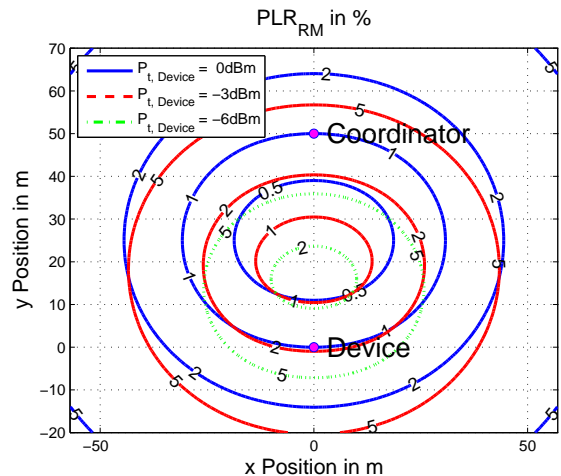}
    \caption{Positions of the relay for RM node to establish $PLR_\mathrm{RM}$ values of $1\%$, $2\%$ and $5\%$ with $PER_\mathrm{D2C}$ of $10\%$ and different transmission power settings of the device $P_\mathrm{t, Device}$.}
    \label{fig_prob_loosing_packets_at_one_prob}
\end{figure}

Of course the reduction of $P_\mathrm{t, Device}$ results in higher data loss, but by placing the relay node according to the probability contours in Fig.~\ref{fig_prob_loosing_packets} the lowest possible $PLR_\mathrm{RM}$ can be achieved. Furthermore, the lower the transmission power of the device, the more sensitive the error probability is against a wrong placement of the relay.

\section{Conclusion}
\label{section_conclusion}
The LLDN mode of the IEEE~802.15.4e standard offers a deterministic behavior for a WSN in star topology where each sensor node transmits up to 100 sensor values per second. The standard already specifies a schedule with a specific time slot for GACK packets and thus the possibility of real-time retransmission within the same superframe. First we proposed the Retransmission Mode, where an additional relay for RM node allows retransmission, if the data packet of the device is lost in the first transmission attempt of the device. The results are lower power consumption in the device and higher transmission reliability. This can be reached with optimized placing of the relay node, which is discussed in this work. The minimum power savings in the device are $33\%$ and can reach up to $50\%$ at lossy channels.

Extended Topology Mode is the second proposal, which enables data transmission from a device, which has only two-hop connection to the coordinator. The power consumption of the device and the relay for ETM are kept low. The cost of this data forwarding is a higher probability of lost data packets.

\section*{Appendix: $PER$ as function of $d$ and $\mathcal{E}_\mathrm{t}$}
\label{appendix_PER_of_d_and_E}

The spatial distribution of the nodes in a  WSN lead to different $PER$s between all involved nodes. We want to derive the $PER$s from one reference link with known $PER_1$. The reference link is characterized by the transmitted bit energy $\mathcal{E}_\mathrm{t,1}$, the distance $d_1$ between transmitter and receiver, and the length of the transmitted packet $l_1$. The second link, for which we want to know the link quality ($PER_2$), is characterized by $d_2$, $\mathcal{E}_\mathrm{t,2}$, and $l_2$. The equations used below are based on a flat Rayleigh fading channel with AWGN. The proposed derivation is performed with only assuming the path loss exponent $\kappa$, which is $2$ for free space propagation and increases in multipath environments.

The first step of the calculation is to derive the bit error rate $BER_\mathrm{1}$ from the known $PER_\mathrm{1}$ of packets with length $l_\mathrm{1}$. We assume independent bit errors and no channel coding:
\begin{equation}
BER_\mathrm{1} = 1-(1-PER_\mathrm{1})^{\frac{1}{l_\mathrm{1}}}.
\label{equ_BER_over_PER}
\end{equation}
The IEEE~802.15.4 PHY uses Offset-QPSK modulation with half-sine pulse shaping, which is equivalent to MSK, in the $2.4\;\mathrm{GHz}$ ISM Band, which has the same bit error probability as QPSK \cite{digital_communication_over_fading_channels}. \cite{fundamentals_wireless_comm} describes the bit error probability of QPSK in a Rayleigh fading channel with AWGN
\begin{equation}
BER = \frac{1}{2}\left( 1-\sqrt{\frac{SNR}{2+SNR}} \right),
\label{equ_BER_over_SNR}
\end{equation}
where $SNR$ describes the signal-to-noise-ratio at the receiver. This yields the known $SNR_\mathrm{1}$ as
\begin{equation}
SNR_\mathrm{1} = \frac{2\left(1-2 BER_\mathrm{1} \right)^2}{1-\left(1-2 BER_\mathrm{1} \right)^2}.
\label{equ_SNR_over_BER}
\end{equation}
In \cite{fundamentals_wireless_comm} $SNR$ is defined as "average received signal energy per symbol time" to "noise energy per symbol time", $SNR:= \frac{\mathcal{E}_\mathrm{r}}{N_0}$. The received bit energy is proportional to $\mathcal{E}_\mathrm{t}/d^{\kappa}$. Thus the relation between known $SNR_\mathrm{1}$ and new $SNR_\mathrm{2}$ can be expressed by
\begin{equation}
\frac{SNR_\mathrm{1}}{SNR_\mathrm{2}} = \frac{\mathcal{E}_\mathrm{t,1} d_2^\kappa }{\mathcal{E}_\mathrm{t,2} d_1^\kappa}
\label{equ_SNR1_to_SN2}
\end{equation}
Thus the $SNR_2$ of the second wireless link can be expressed by the known $SNR_\mathrm{1}$, without knowledge about noise, antenna gain, etc. From (\ref{equ_BER_over_SNR}) and (\ref{equ_SNR1_to_SN2}) we find
\begin{equation}
BER_2 = \frac{1}{2}\left( 1-\sqrt{\frac{1}{1+\left(2 \mathcal{E}_\mathrm{t,1} d_2^\kappa\right)/ \left( \mathcal{E}_\mathrm{t,2} d_1^\kappa SNR_1\right)}} \right),
\label{equ_BER2_over_SNR1}
\end{equation}
which finally leads to
\begin{equation}
PER_2 = 1-\left(1-\frac{1}{2}\left( 1-\sqrt{\frac{1}{1+ \frac{2 \mathcal{E}_\mathrm{t,1} d_2^\kappa}{\mathcal{E}_\mathrm{t,2} d_1^\kappa SNR_1} }} \right)\right)^{l_2}.
\label{equ_PER2_over_SNR1}
\end{equation}
\nocite{ex1,ex2}
\bibliography{my_bib}

\begin{thebibliography}{10}
\providecommand{\url}[1]{#1}
\csname url@samestyle\endcsname
\providecommand{\newblock}{\relax}
\providecommand{\bibinfo}[2]{#2}
\providecommand{\BIBentrySTDinterwordspacing}{\spaceskip=0pt\relax}
\providecommand{\BIBentryALTinterwordstretchfactor}{4}
\providecommand{\BIBentryALTinterwordspacing}{\spaceskip=\fontdimen2\font plus
\BIBentryALTinterwordstretchfactor\fontdimen3\font minus
  \fontdimen4\font\relax}
\providecommand{\BIBforeignlanguage}[2]{{%
\expandafter\ifx\csname l@#1\endcsname\relax
\typeout{** WARNING: IEEEtran.bst: No hyphenation pattern has been}%
\typeout{** loaded for the language `#1'. Using the pattern for}%
\typeout{** the default language instead.}%
\else
\language=\csname l@#1\endcsname
\fi
#2}}
\providecommand{\BIBdecl}{\relax}
\BIBdecl

\bibitem{industrial_WSN_challenges}
V.~Gungor and G.~Hancke, ``Industrial wireless sensor networks: Challenges,
  design principles, and technical approaches,'' \emph{Industrial Electronics,
  IEEE Transactions on}, vol.~56, no.~10, pp. 4258--4265, 2009.

\bibitem{energy_harvesting_technologies}
U.~Alvarado, A.~Juanicorena, I.~Adin, B.~Sedano, I.~Guti\'{e}rrez, and
  J.~de~N\'{o}, ``Energy harvesting technologies for low-power electronics,''
  \emph{Transactions on Emerging Telecommunications Technologies}, vol.~23,
  no.~8, pp. 728--741.

\bibitem{IEEE_STD_802.15.4}
``Wireless medium access control {(MAC)} and physical layer {(PHY)}
  specifications for low-rate wireless personal area networks {(WPANs)},''
  \emph{IEEE Std 802.15.4-2006 (Revision of IEEE Std 802.15.4-2003)}, 2006.

\bibitem{CC2520}
\BIBentryALTinterwordspacing
(2011) {CC2520: 2.4 GHz IEEE 802.15.4/ZigBee RF Tranceiver}. Texas Instruments.
  [Online]. Available: \url{http://www.ti.com/lit/gpn/cc2520}
\BIBentrySTDinterwordspacing

\bibitem{MRF24J40}
\BIBentryALTinterwordspacing
(2010) {MRF24J40: IEEE 802.15.4 2.4 GHz RF Transceiver}. Microchip Technology
  Inc. [Online]. Available:
  \url{http://ww1.microchip.com/downloads/en/DeviceDoc/39776C.pdf}
\BIBentrySTDinterwordspacing

\bibitem{wirelessHART_versus_ISA100_11a}
S.~Petersen and S.~Carlsen, ``Wireless{HART} versus {ISA}100.11a: The format
  war hits the factory floor,'' \emph{Industrial Electronics Magazine, IEEE},
  vol.~5, no.~4, pp. 23 --34, Dec. 2011.

\bibitem{wirelessHART_energy_analysis}
O.~Khader, A.~Willig, and A.~Wolisz, ``Wireless{HART} {TDMA} protocol
  performance evaluation using response surface methodology,'' in
  \emph{Broadband and Wireless Computing, Communication and Applications
  (BWCCA), 2011 International Conference on}, Oct. 2011, pp. 197 --206.

\bibitem{guarantee_real_time_in_802_15_4}
S.-E. Yoo, P.~K. Chong, D.~Kim, Y.~Doh, M.-L. Pham, E.~Choi, and J.~Huh,
  ``Guaranteeing real-time services for industrial wireless sensor networks
  with {IEEE} 802.15.4,'' \emph{Industrial Electronics, IEEE Transactions on},
  vol.~57, no.~11, pp. 3868--3876, 2010.

\bibitem{IEEE_STD_802.15.4e-2012}
``{IEEE} standard for local and metropolitan area networks -- part 15.4:
  Low-rate wireless personal area networks {(LR-WPANs) Amendment 1: {MAC}
  sublayer},'' \emph{IEEE Std 802.15.4e-2012 (Amendment to IEEE Std
  802.15.4-2011)}, 2012.

\bibitem{first_variant_of_IEEE_802.15.4e}
F.~Chen, T.~Talanis, R.~German, and F.~Dressler, ``Real-time enabled {IEEE}
  802.15.4 sensor networks in industrial automation,'' in \emph{Industrial
  Embedded Systems, 2009. SIES '09. IEEE International Symposium on}, July
  2009, pp. 136 --139.

\bibitem{study_of_IEEE_802.15.4e}
F.~Chen, R.~German, and F.~Dressler, ``Towards {IEEE} 802.15.4e: A study of
  performance aspects,'' in \emph{Pervasive Computing and Communications
  Workshops (PERCOM Workshops), 2010 8th IEEE International Conference on},
  April 2010, pp. 68 --73.

\bibitem{optimal_position_in_two_tiered_WSN}
R.~Tripathi, Y.~Singh, and N.~Verma, ``Two-tiered wireless sensor networks -
  base station optimal positioning case study,'' \emph{Wireless Sensor Systems,
  IET}, vol.~2, no.~4, pp. 351--360, 2012.

\bibitem{survey_multipoint_relay}
O.~Liang, Y.~A. Sekercioglu, and N.~Mani, ``A survey of multipoint relay based
  broadcast schemes in wireless ad hoc networks,'' \emph{Commun. Surveys
  Tuts.}, vol.~8, no.~4, pp. 30--46, Oct. 2006.

\bibitem{cross_layer_topology_optimization}
F.~Cuomo, A.~Abbagnale, and E.~Cipollone, ``Cross-layer network formation for
  energy-efficient {IEEE} 802.15.4/zigbee wireless sensor networks,'' \emph{Ad
  Hoc Networks}, vol.~11, no.~2, pp. 672 -- 686, 2013, special Issue on
  Cross-layer design in ad hoc and sensor networks.

\bibitem{routing_in_wsn}
J.~Al-Karaki and A.~Kamal, ``Routing techniques in wireless sensor networks: a
  survey,'' \emph{Wireless Communications, IEEE}, vol.~11, no.~6, pp. 6 -- 28,
  Dec. 2004.

\bibitem{TSMP}
K.~S.~J. Pister and L.~Doherty, ``{TSMP}: Time synchronized mesh protocol,'' in
  \emph{In Proc. of the IASTED Intern. Symposium on Distributed Sensor Networks
  (DSN08)}, 2008.

\bibitem{Adapt_Sync_in_802_15_4e}
D.~Stanislowski, X.~Vilajosana, Q.~Wang, T.~Watteyne, and K.~Pister, ``Adaptive
  synchronization in {IEEE} 802.15.4e networks,'' \emph{Industrial Informatics,
  IEEE Transactions on}, vol.~PP, no.~99, pp. 1--1, 2013.

\bibitem{cooperative_diversity_in_WLAN_and_WSN}
R.~Khan and H.~Karl, ``Mac protocols for cooperative diversity in wireless lans
  and wireless sensor networks,'' \emph{Communications Surveys Tutorials,
  IEEE}, vol.~16, no.~1, pp. 46--63, First 2014.

\bibitem{cooperative_protocols_survey}
P.~Ju, W.~Song, and D.~Zhou, ``Survey on cooperative medium access control
  protocols,'' \emph{Communications, IET}, vol.~7, no.~9, pp. 893--902, June
  2013.

\bibitem{performance_of_relay_selection}
Al-Tous and Barhumi, ``Performance analysis of relay selection in cooperative
  networks over rayleigh flat fading channels,'' \emph{EURASIP Journal on
  Wireless Communications and Networking}, 2012.

\bibitem{cooperative_diversity}
J.~Laneman, D.~Tse, and G.~W. Wornell, ``Cooperative diversity in wireless
  networks: Efficient protocols and outage behavior,'' \emph{Information
  Theory, {IEEE} Transactions on}, vol.~50, no.~12, pp. 3062--3080, 2004.

\bibitem{srinivasan_kappa_factor}
K.~Srinivasan, M.~Jain, J.~I. Choi, T.~Azim, E.~S. Kim, P.~Levis, and
  B.~Krishnamachari, ``The $\kappa$ factor: inferring protocol performance
  using inter-link reception correlation,'' in \emph{Proceedings of the
  sixteenth annual international conference on Mobile computing and
  networking}, ser. MobiCom '10.\hskip 1em plus 0.5em minus 0.4em\relax New
  York, NY, USA: ACM, 2010, pp. 317--328.

\bibitem{srinivasan_beta_factor}
K.~Srinivasan, M.~A. Kazandjieva, S.~Agarwal, and P.~Levis, ``The $\beta$
  factor: measuring wireless link burstiness,'' in \emph{SenSys '08:
  Proceedings of the 6th ACM conference on Embedded network sensor
  systems}.\hskip 1em plus 0.5em minus 0.4em\relax New York, NY, USA: ACM,
  2008, pp. 29--42.

\bibitem{fundamentals_wireless_comm}
D.~Tse and P.~Viswanath, \emph{Fundamentals of Wireless Communication}.\hskip
  1em plus 0.5em minus 0.4em\relax NY, USA: Cambridge University Press, 2005.

\bibitem{channel_measurements_koerber_scholl}
H.-J. K\"orber, H.~Wattar, and G.~Scholl, ``Modular wireless real-time
  sensor/actuator network for factory automation applications,''
  \emph{Industrial Informatics, IEEE Transactions on}, vol.~3, no.~2, pp.
  111--119, May 2007.

\bibitem{PER_for_rotating_sensor_nodes}
G.~Anastasi, M.~Conti, and M.~Di~Francesco, ``A comprehensive analysis of the
  {MAC} unreliability problem in {IEEE} 802.15.4 wireless sensor networks,''
  \emph{Industrial Informatics, IEEE Transactions on}, vol.~7, no.~1, pp.
  52--65, 2011.

\bibitem{speed_dependent_PER}
L.~Tang, K.-C. Wang, and Y.~Huang, ``Study of speed-dependent packet error rate
  for wireless sensor on rotating mechanical structures,'' \emph{Industrial
  Informatics, IEEE Transactions on}, vol.~9, no.~1, pp. 72--80, 2013.

\bibitem{RAPP}
T.~Rappaport, \emph{Wireless Communications: Principles and Practice},
  2nd~ed.\hskip 1em plus 0.5em minus 0.4em\relax Upper Saddle River, NJ, USA:
  Prentice Hall PTR, 2001.

\bibitem{channel_measurement_for_802_15_4_transceiver}
L.~Tang, K.-C. Wang, Y.~Huang, and F.~Gu, ``Channel characterization and link
  quality assessment of {IEEE} 802.15.4-compliant radio for factory
  environments,'' \emph{Industrial Informatics, IEEE Transactions on}, vol.~3,
  no.~2, pp. 99--110, 2007.

\bibitem{XORs_in_the_air}
S.~Katti, H.~Rahul, W.~Hu, D.~Katabi, M.~Medard, and J.~Crowcroft, ``{XORs} in
  the air: Practical wireless network coding,'' \emph{Networking, IEEE/ACM
  Transactions on}, vol.~16, no.~3, pp. 497 --510, June 2008.

\bibitem{TDMA_for_testbed}
A.~Berger, A.~P\"otsch, and A.~Springer, ``{TDMA} approach for efficient data
  collection in wireless sensor networks,'' in \emph{Emerging Technologies and
  Factory Automation, 2012. ETFA 2012. IEEE International Conference on},
  September 2012.

\bibitem{digital_communication_over_fading_channels}
M.~Simon and M.~Alouini, \emph{Digital Communication over Fading Channels: A
  Unified Approach to Performance Analysis}, ser. Wiley Series in
  Telecommunications and Signal Processing.\hskip 1em plus 0.5em minus
  0.4em\relax NY, USA: John Wiley \& Sons, 2000.

\end{thebibliography}
\bibliographystyle{IEEEtran}

\end{document}